\documentclass[preprint,showkeys,aps,prb]{revtex4}
\usepackage[dvips]{graphicx}

\begin{document}

\title{Yokohama to Ruby Valley : Around the World in 80 Years. II.}

\author{
Carol Griswold Hoover \& William Graham Hoover  \\
Ruby Valley Research Institute                  \\
Highway Contract 60, Box 601                    \\
Ruby Valley, Nevada 89833                       \\
}

\date{\today}

\keywords
{Molecular Dynamics, SPAM, Lyapunov Instability, Time-Reversible Thermostats, Chaos}

\vspace{0.1cm}

\begin{abstract}
We two had year-long research leaves in Japan, working together fulltime with several Japanese plus
Tony De Groot back in Livermore and Harald Posch in Vienna. We summarize a few of the high spots from
that very productive year ( 1989-1990 ), followed by an additional fifteen years' work in Livermore,
with extensive  travel.  Next came our retirement in Nevada in 2005, which has turned out to be a
long-term working vacation.  Carol narrates this part of our history together.
\end{abstract}

\maketitle

\section{To Yokohama from Livermore}

I first met Bill in 1972 as a student in his graduate courses in Statistical Mechanics and Kinetic Theory.
I had the welcome opportunity to work part time at the Lawrence Livermore National Laboratory (LLNL) as a
``student employee'' while enrolled in the PhD program at the Department of Applied Science (DAS) in the
University of California at Davis-Livermore. Although I had selected a talented thesis advisor, John
Killeen, who specialized in computational plasma physics, I recall thinking at the time that I would have
enjoyed thesis research with Bill.  His classes were both stimulating and challenging!

Sixteen years later I was a Group Leader for the National Magnetic Fusion Energy Computer Center (NMFECC) where
innovations in computer hardware and software led to the first interactive operating system, to the first
unclassified national network, and to the era of the Cray supercomputers. I taught scientists and engineers the
implementation and use of algorithms on computers still in the design phase.  By then my interests had broadened
to include finite-element simulations.

I helped many plasma physicists to vectorize their software.  One of them, a Professor Sumnesh Gupta from Lousiana State
University, asked me to arrange a meeting with Bill during his visit.  Bill is and was decisive. When I went to
check on his meeting with Professor Gupta he asked me ``How about dinner?''  My reply was, ``Are you serious ?'' !
That was the beginning of our wonderful life-long partnership.  We were married less than a year later.

In 1988 we visited Cornell.  Bill met with Ben Widom and his chemistry colleagues while I met with a group
specializing in transputer-based parallel systems.  The Crays were becoming prohibitively expensive and were
nearing their speed limits. What we learned at Cornell was sufficiently exciting that on our return to Livermore
we contacted another DAS PhD, Tony De Groot.  His thesis research analyzed arrays of interconnected low-cost
transputers with a fast network, providing a parallel architecture alternative to the vector parallelism used
in the Cray supercomputers.  Tony had a grant to build a prototype system with 64 processors and to demonstrate
its speed for an interesting physics or engineering problem. {\bf Figure 1} shows a snapshot from the
project. Tony's work was a perfect match for our interests in a future with parallel computers, molecular dynamics,
materials science, and continuum mechanics all linked together.

\begin{figure}
\includegraphics[width=3.5in]{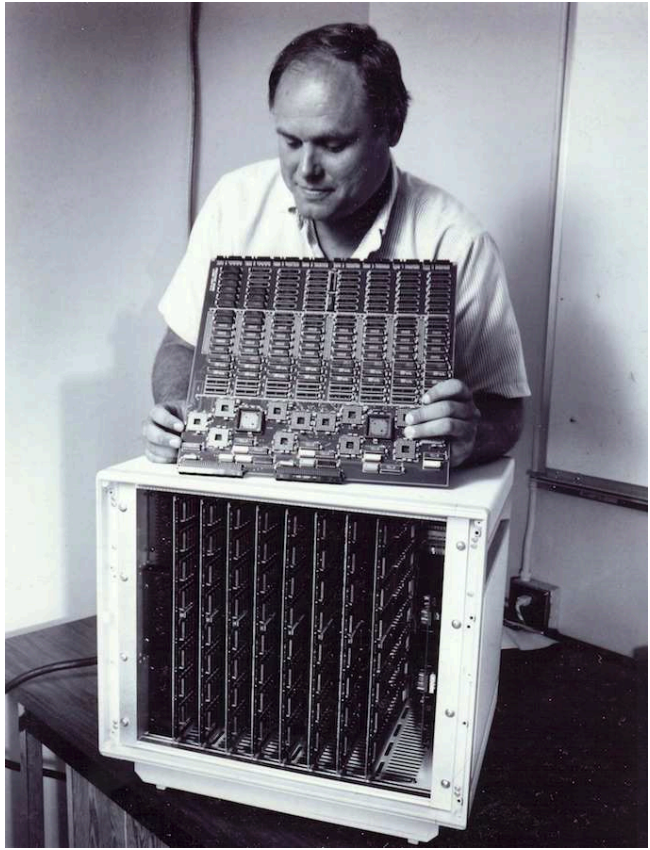}
\caption{
( {\bf S}ystolic {\bf P}rocessor with {\bf R}econfigurable {\bf I}nterconnection {\bf N}etwork of {\bf T}ransputers ).
}
\end{figure}

\begin{figure}
\includegraphics[width=4.5in,angle=-90.]{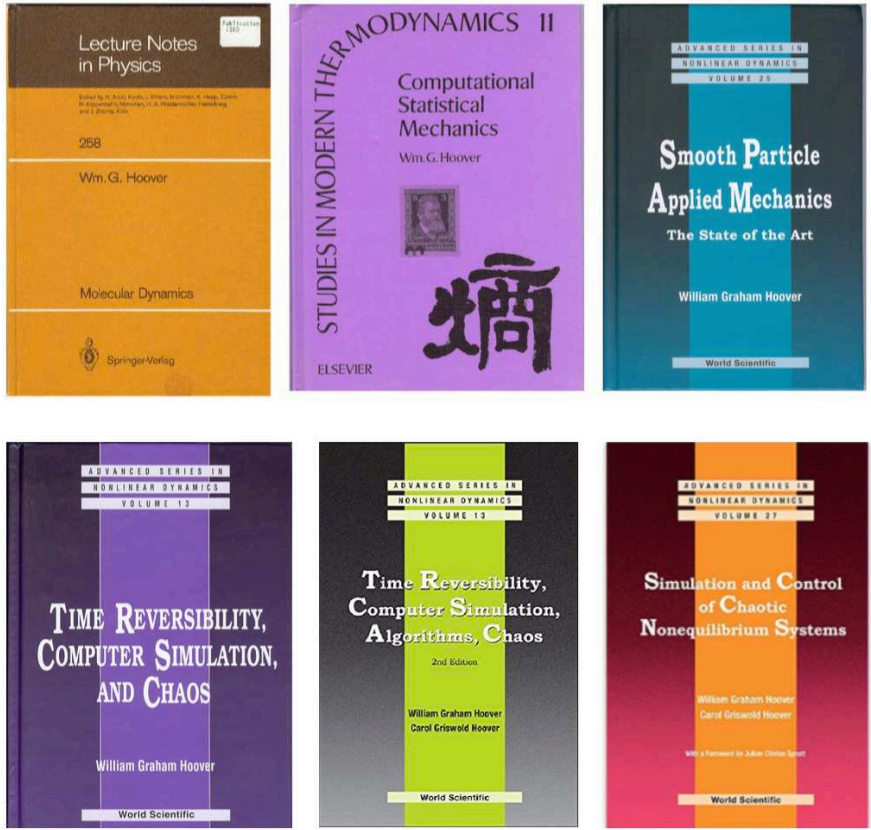}
\caption{
Bill has written six books.  He completed the second book, {\it Computational Statistical Mechanics}, while
we were in Japan.  Harald Posch gave Bill the Boltzmann stamp and Shuichi Nos\'e furnished the kanji on the
cover. The two-part kanji ``entropy'' combines those for  ``heat'' and ``divide''.
}
\end{figure}

\subsection{Research Leave, Sabbatical, and a Collaboration in Japan, 1989-1990 }

In 1989 it was time for Bill's third Sabbatical Leave from DAS.  Bill was invited to Japan to visit Shuichi Nos\'e
at Keio University in Yokahama. I was invited to work with Professor Toshio Kawai, who headed up the same physics
department.  We both had supporting grants, Bill's from the Japan Society for the Promotion of Science and mine a
research and teaching grant from LLNL.

We spent the academic year 1989-1990 in Nos\'e's Keio University laboratory. Officially I was working with Professor
Kawai, who had an interest in statistical mechanics while Bill was working with Shuichi Nos\'e.  But in reality we both
worked together with Professor Taisuke Boku, a parallel-computing specialist, and a physicist Sigeo Ihara at Hitachi's
Kokubunji Laboratory. Our common interest was large-scale molecular dynamics problems involving plastic flow. This work
was coordinated with several colleagues back at Livermore as well as Brad Holian at Los Alamos.  Tony De Groot made the
project possible.  He successfully designed and built the SPRINT computer which he could run fulltime, at Cray speed,
in his office.

We were also collaborating with Harald Posch in Wien on Lyapunov spectra for many-body chains
and pendula\cite{b1} while writing a book on Computational Statistical mechanics\cite{b2}. See
{\bf Figure 2}. Bill's Son Nathan, who had married us back in Livermore, was working at Teradyne
Tokyo, and accompanied by his Wife Megumi, throughout our research leaves in nearby Yokohama, leading
to many Family get-togethers, often with our Japanese colleagues and their Families.


\begin{figure}
\includegraphics[width=5in,angle=-90.]{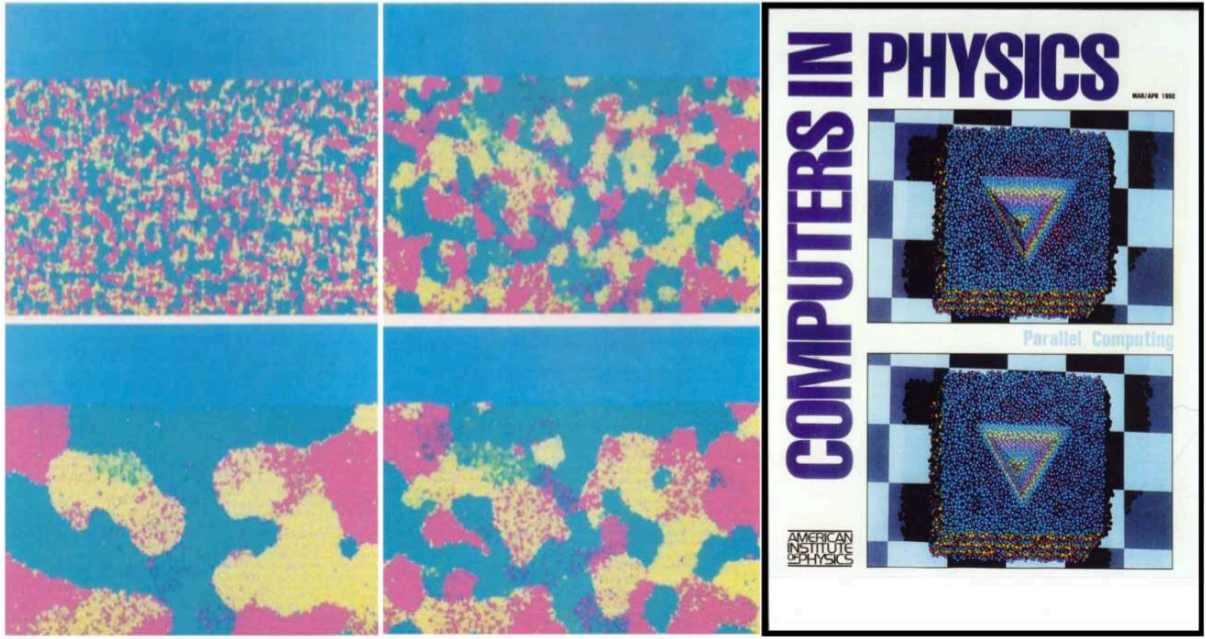}
\caption{
The left panel shows the annealing of an initially unstable square lattice using a combined Lennard-Jones
and embedded-atom potential.  This sample was used as the initial condition for modelling the
indentation of copper.  We simulated the indentation of silicon in three dimensions with both atomistic
and planar-surface indentors. The speed of the million-atom simulation on the 64-processor SPRINT
equalled the speed of a Cray-2.  The picture shown on the right appeared on the cover of Computers
in Physics in 1992 and illustrates the indentation of 373,248 silicon atoms.
}
\end{figure}

\subsection{Large-Scale Molecular Dynamics Simulations -- ``Plastic Flow''}

Our biggest project in Japan involved a Baker's dozen of physicists, engineers, and computer
scientists. The goal was to model, visualize, and analyze  million-atom two- and three-dimensional
``plastic flows'', flows of
solids in response to shear stress.  The two-dimensional work allowed us to study the kinetics of
grain formation during annealing.  See {\bf Figure 3}. The three-dimensional work modeled the
indentation of silicon and ductile metals using angle-dependent and embedded-atom force models in
the dynamics. All of this many-body work was carried out on Tony's SPRINT computer.  As fast as a
CRAY, SPRINT was put together with \$30,000 worth of transputers rather than \$30,000,000 of CRAY
hardware. This collaboration resulted in several papers over a six-year period, one of them with
nine authors, three in Japan, three in Livermore, Brad at Los Alamos, plus Bill and me\cite{b3,b4,b5}.
Late at night in the Kawai Lab Bill was writing his {\it Computational Statistical Mechanics} book
with some help from me.

\section{Fifteen Years at the Livermore Laboratory, 1990-2005}
\begin{figure}
\includegraphics[width=5in,angle=+90]{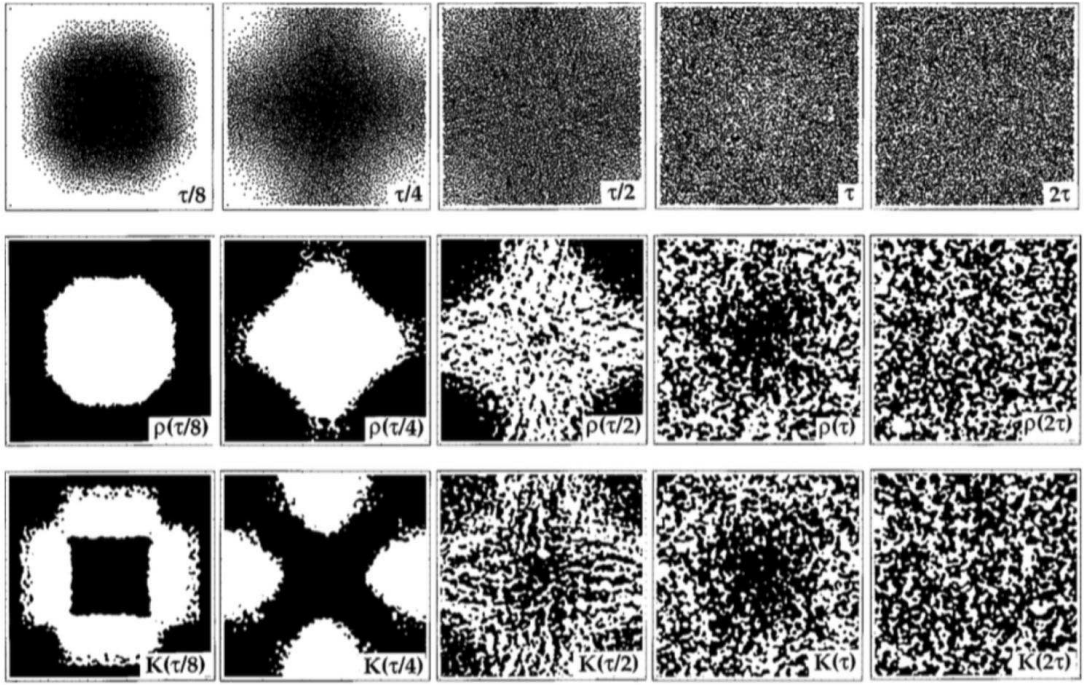}
\caption{
Snapshots of a 16,384 particle simulation of the four-fold adiabatic expansion of an ideal gas, using Lucy's
weight function with $h = 6$.  The individual particle locations, as well as grid-based contour
representations of the density and kinetic energy, are shown at times of
$(\tau/8),(\tau/4),(\tau/2),(\tau),(2\tau)$ where $\tau$ is the sound traversal time.
The contour plots of the density ( middle row ) and the kinetic energy ( bottom row ) show above-average regions
in white and below-average in black\cite{b14}. The density change in expansion is from 1 to 0.25 .
}
\end{figure}

The background of the 1980s was foundational for our research at Livermore : Howard Hanley's
1982 Boulder Conference ``Nonlinear Fluid Behavior''\cite{b6}; Nos\'e's two 1984
papers\cite{b7,b8}; Bill's followup ``Nos\'e-Hoover'' paper\cite{b9}; and Giovanni Ciccotti and Bill's
1985 Fermi School\cite{b10} ``Molecular-Dynamics Simulation of Statistical-Mechanical Systems''.
These beginnings soon led to the discovery of the ubiquitous fractal distributions stemming from
time-reversible simulations consistent with the Second Law of Thermodynamics\cite{b11}. Bill and I,
along with many of those in this room, as well as our hundreds of colleagues, have explored and
analyzed thousands of different model systems so as to flesh out the connection of time-reversible
microscopic mechanics with irreversible macroscopic thermodynamics and computational fluid dynamics.

The goal for this period was a consistent and comprehensive view of macroscopic and microscopic
nonequilibrium systems. These were years of exploration and learning, applying dynamical-systems
methods to small and large atomistic simulations. Gibbs' phase-space description, augmented by
feedback and constraints, relates entropy change to the action of thermostats extracting the
irreversible heat generated by velocity or temperature gradients.  Here we review a few examples.
We consider macroscopic nonequilibria first, taking up an algorithm resembling {\it molecular}
dynamics but developed in order to solve {\it continuum} problems, Smooth-Particle Applied
Mechanics, SPAM.

\subsection{The Structure of Smooth-Particle Applied Mechanics = SPAM}

SPAM was developed in 1977 by Lucy and by Monaghan and Gingold at Cambridge\cite{b12,b13}.  They
solved large-scale astrophysical problems by ascribing all of the many {\it continuum} properties
( density, velocity, energy, stress, heat flux . . . ) to a mesh-free set of moving {\it particles}.
This same idea can be applied to traditional problems in computational fluid dynamics such
as shockwave structure and Rayleigh-B\'enard convection.  It can also be merged with molecular
dynamics, defining an equivalent continuum within which the moving particles serve as
interpolation points.  The key idea is the smooth-particle interpolation scheme
: localized {\it particle} properties are patched together to generate global very smooth 
doubly-differentiable {\it continuum fields}. These fields are constructed by summing particle
contributions with a weight function which has two continuous derivatives at the cutoff distance $h$ :
$$
w_{\rm Lucy}(r<h) = C_D [ \ 1 + 3(r/h) \ ] [ \ 1 - (r/h) \ ]^3 \ .
$$
The range $h$ of the weight function is chosen to include a few dozen particles.  For
simplicity here we choose to use particles of unit mass. The normalization constant $C_D$
depends upon the dimensionality $D$ of the problem :
$$
C_1 = (5/4h) \ ; \ C_2 = (5/\pi h^2) \ ; \ C_3 = (105/16\pi h^3) \longleftrightarrow
\int w{\rm dV} \equiv 1 \ .                                    
$$
Though ( many ) other weight functions can be used Lucy's is the simplest polynomial that
satisfies the constraints of a smooth maximum with two vanishing derivatives at the maximum
distance $h$ .


\begin{figure}
\includegraphics[width=4.5in,angle=+90]{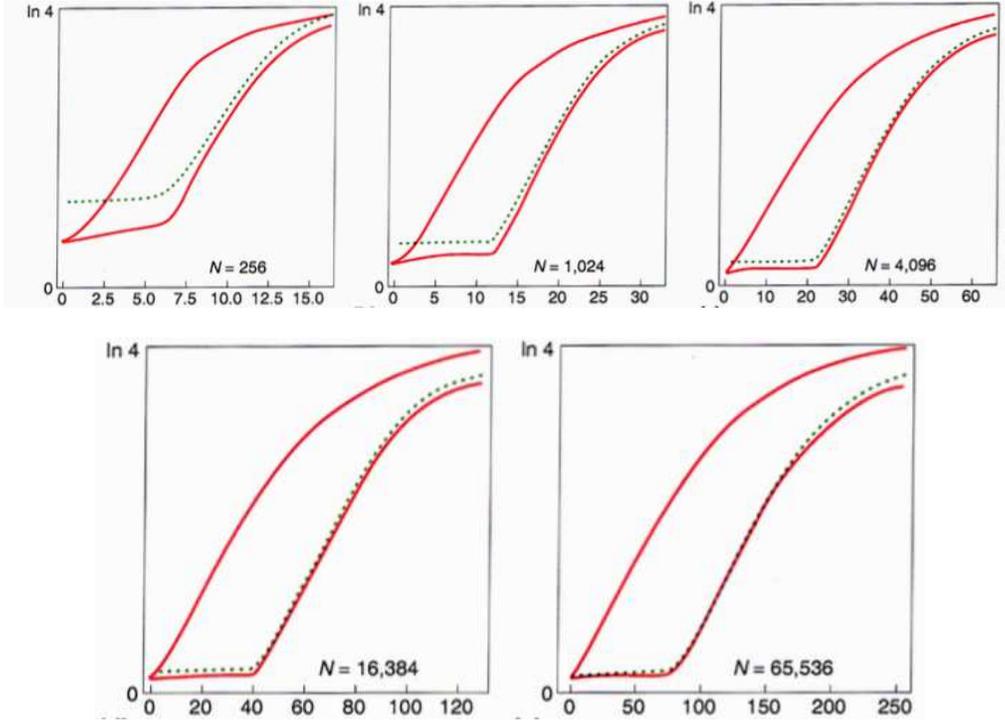}
\caption{
Time development of the ``laboratory-frame'' entropy ( upper curve ), ``comoving'' ``Lagrangian-frame''
entropy ( lower curve ), as well as a fixed-grid-based version of the comoving entropy ( dots ) for
systems of 256, 1024, 4096, 16384, and 65536 particles.  The time interval shown is one-half the sound
traversal time in each case.  The entropy range shown is $Nk\ln(4)$ in each case.
}
\end{figure}

Fluid simulations solve ordinary differential equations for the three particle properties
$\{ \ r_i,v_i,e_i \ \}$ . From these and the gradients which can be derived from
them by differentiation all the continuum properties
$\{ \ \rho(r), u(r), e(r),\sigma(r),Q(r), . . . \ \}$ follow :
$$
\rho(r) = \sum_i w(r-r_i) \ ; \ \rho(r)u(r) = \sum_i w(r-r_i)v_i \ ;
\ \rho(r)e(r) = \sum_i w(r-r_i)e_i \ .
$$

The continuity equation ,
$$
(\partial \rho/\partial t) = -\nabla \cdot (\rho u) \longleftrightarrow
\dot \rho = (\partial \rho/\partial t) + u\cdot \nabla \rho = -\rho \nabla \cdot u \ ,
$$
is an identity with this approach. It is solved {\it automatically} by summing weight
functions, leaving only the equations for $( \ \dot r, \ \dot v,  \ \dot e \ )$ to
solve for each particle.
For a two-dimensional ideal gas, with $P = (\rho^2/2) = \rho e$ , the equation of
motion is familiar :
$$
\ddot r_i = \dot v_i = -\sum_j[ \ (P/\rho^2)_i + (P/\rho^2)_j \ ]\nabla_iw_{ij} =
-\sum_j \nabla_iw_{ij} \ .
$$
These motion equations for the smooth particles are {\it identical}, isomorphic
to those for a gas with the weak repulsive pairwise-additive potential $w(r)$ .

As a simple example of the applicability of SPAM to molecular dynamics, we consider
the free expansion problem illustrated in {\bf Figures 4 and 5}. As the motion
proceeds, governed by $w(r)$ , this same smooth-particle weighting function can be
used to define and measure the local values of the density, internal energy, kinetic
energy, and ( local-equilibrium ) entropy fields of the expanding and equilibrating
fluid. Let us demonstrate SPAM by applying it to a difficult problem, the four-fold
adiabatic free expansion of a compressed gas into a closed periodic container.

\subsection{Free Expansion with SPAM}
\begin{figure}
\includegraphics[width=4in,angle=-90]{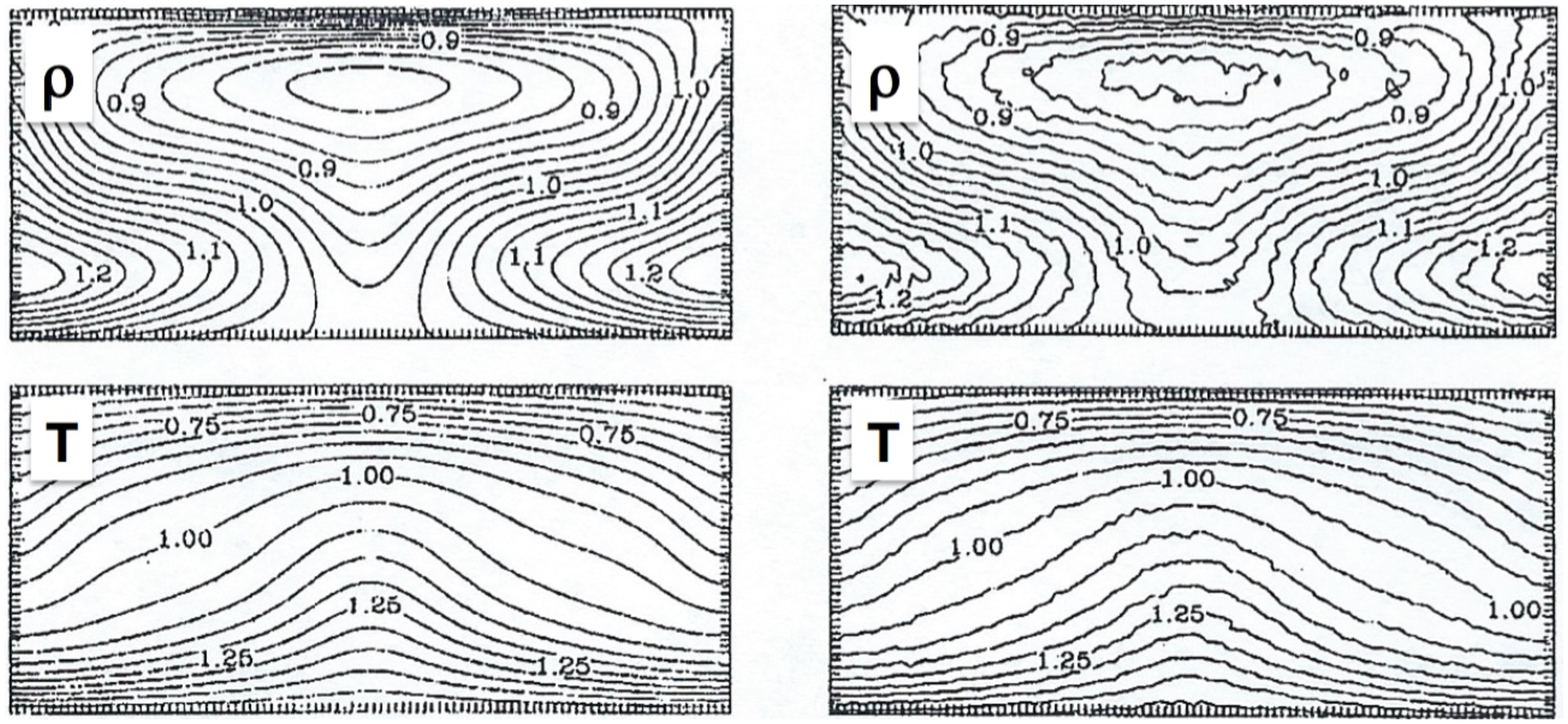}
\caption{
Rayleigh-B\'enard rolls simulated with finite differences (left) and 
with SPAM (right) for a Rayleigh number of 10000. Upper panels are density contours
and the lower panels are temperature contours.  The temperatures at the top and 
bottom are respectively, 0.5 and 1.5 . The overall density is 1.0.  The thermal 
diffusivity and kinematic viscosity are both chosen equal to 0.4 .
}
\end{figure}

\begin{figure}
\includegraphics[width=5.5in,angle=-90]{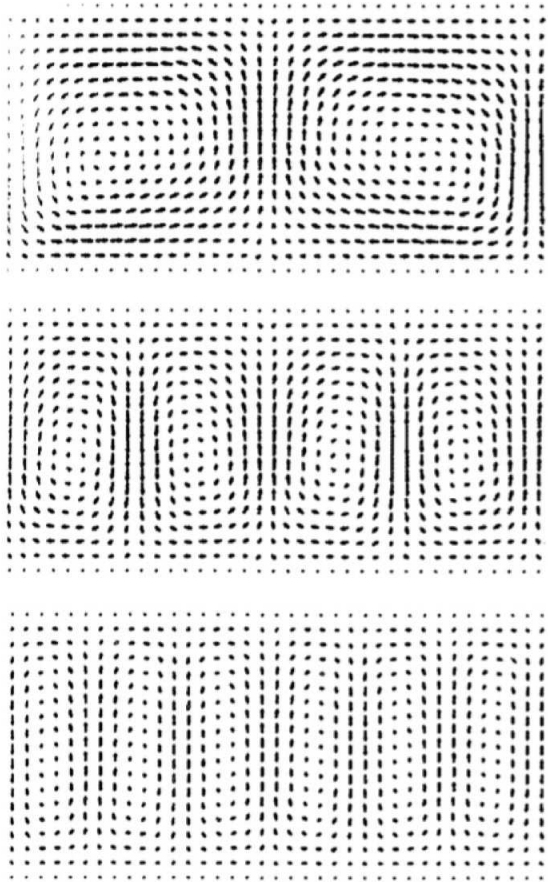}
\caption{
 Two-roll, four-roll, and six-roll stationary flows for a two-dimensional ideal gas 
with $R = 40,000$ and Prandtl number unity.  Fully stationary two-roll solutions and
four-roll solutions were obtained for meshes with 36x18, 72x36 and 144x72 nodes.  The
larger meshes show also six-roll solutions.
}
\end{figure}

\begin{figure}
\includegraphics[width=5.5in,angle=-90]{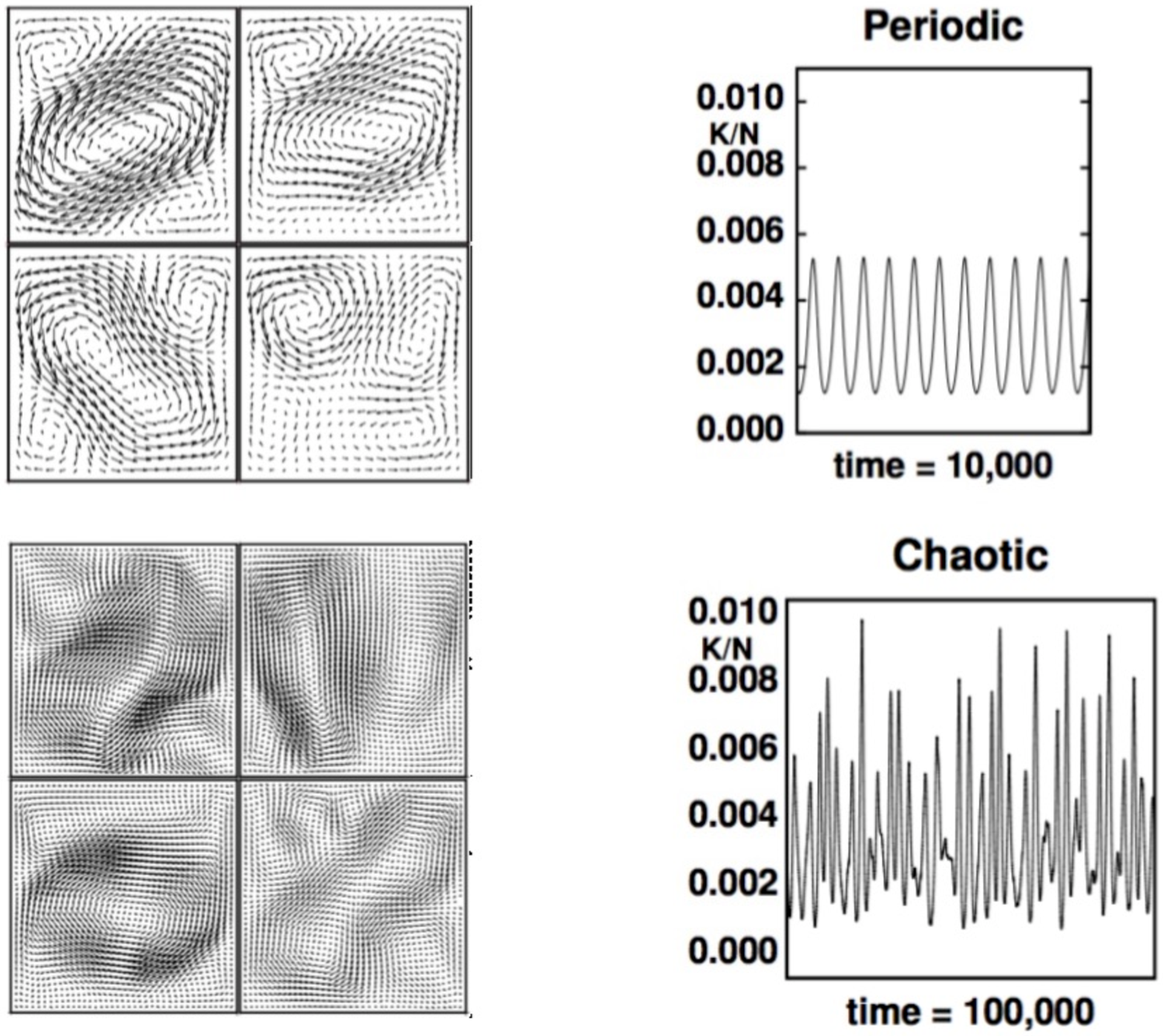}
\caption{
Morphology changes for high Rayleigh numbers.  
Harmonic morphology ( Top left ) of Rayleigh-B\'enard rolls for a Rayleigh number of 160000. 
This solution uses an $ 80\times 80 $ computational grid . Times shown are at one-eight of a
perod.  Chaotic morphology ( Bottom left ) for a Rayleigh number of 800000 using
a $160 \times 160$ computational grid. Time interval between the snapshots is 1000. 
The corresponding kinetic energies per particle are shown on the right side of the figure.
See Section 4.11.2 of the fifth book of six shown in {\bf Figure 2}.
}
\end{figure}

With Harald and Vic Castillo\cite{b14,b15} we used the motion-equation isomorphism
to study the expansion of an ideal-gas fluid into a periodic volume four times larger
than the original.  See {\bf Figure 4}.  Our goal was better to understand the
entropy increase associated with the expansion process, $Nk\ln(4)$ , where $k$ is
Boltzmann's constant.  The ( equilibrium only ! ) entropy for the two-dimensional
ideal gas is just $S = Nk\ln(e/\rho)$ .

{\bf Figure 4} shows five stages in the expansion process.  At the top we see the
particle locations.  Below are the density and the kinetic energy.  The white
regions are above average and the black below.  Three very significant points of interest were
clarified by these simulations : [ 1 ] Equilibrium occurs quickly, in a time
on the order of a sound traversal time; [ 2 ] The kinetic energy field, which
contributes to the local energy and entropy, needs to be measured {\it relative} to the
local stream velocity; and [ 3 ] Entropy {\it must} be evaluated using the local energy
( including the {\it local} fluctuating kinetic energy ) rather than the laboratory-frame
energy.  This last point becomes clear on looking at the time dependence of the two
kinetic-energy calculations.  Entropy increase ( in isentropic expansions ) doesn't
occur until the gas contacts its periodic images and begins the dissipation of mass
motion into heat.  See {\bf Figure 5} .  The free expansion problem\cite{b14,b15}
is particularly challenging for finite-element approaches due to its severe shear
deformation and demonstrates the ability of SPAM to deal with interesting fluid
mechanics problems.  The simplicity of SPAM and its computer implementation make
it an ideal teaching tool.

\subsection{Rayleigh-B\'enard Convection with SPAM}

Another SPAM application is Rayleigh-B\'enard convection\cite{b16,b17,b18}. A
fluid heated from below and subject to a gravitational field can exhibit several
morphologies, depending upon the dimensionless Rayleigh Number. Again we choose a
two-dimensional ideal gas. Here the system height is $H$ and the gravitational
field strength is chosen consistent with constant density, $g = (k\Delta T/H)$ .
The top-to-bottom temperature difference $\Delta T$ is equal to the mean temperature.
The Rayleigh Number is ${\cal R} = (gH^3/\nu D) = 10000$ . Here $\nu$ and $D$ are the
kinematic viscosity and the thermal diffusivity.  In Oyeon Kum's PhD work he compared
simulations with straightforward-but-tedious Eulerian continuum mechanics to SPAM.
See {\bf Figure 6} for a two-roll solution which is stable at ${\cal R} = 
10000$ .  At a higher value of the Rayleigh number, 40000 , three different
stationary solutions of the continuum equations occur\cite{b17,b18}.  See
{\bf Figure 7}. 

Though maximum entropy or minimum entropy production have been touted
as useful concepts in defining stability these examples show that there is no
meaningful way to distinguish the relative stability ( as in thermodynamic phase
stability ) of the different roll patterns.  As the Rayleigh Number is increased
through 90,000 the rolls begin regular vertical ``harmonic'' oscillations.  Between
150,000 and 250,000 there are two separate solutions of the motion equations, one
regular and harmonic, the other chaotic. See {\bf Figure 8}. The harmonic solution
transports heat a bit more efficiently than does its chaotic brother\cite{b18}.

\subsection{Lyapunov Instability and the Second Law of Thermodynamics}

\begin{figure}
\includegraphics[width=5in,angle=-90]{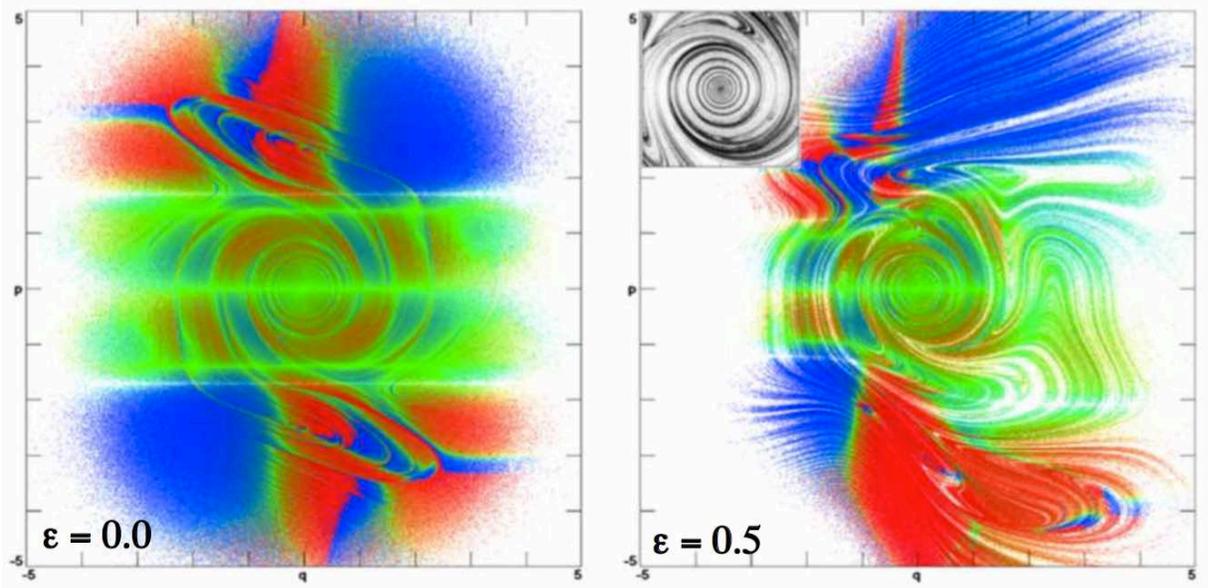}
\caption{
Penetrations of the $( \ q,p,0 \ )$ plane for the chaotic and ergodic 0532 Model using an adaptive
fourth-order Runge-Kutta integration with a timestep $dt \simeq 0.001$ .  The red and blue points
correspond to maximum and minimum values of the local Lyapunov exponent.  The equilibrium 
$( \ \zeta =  0 \ )$ cross section at the left shows inversion symmetry, corresponding to viewing the
oscillator in a mirror.  The lack of symmetry about the horizontal $( \ p = 0 \ )$ axis shows that the 
exponents depend upon the past rather than the future.  The nonequilibrium section ( $\epsilon = 0.50$ )
shown to the right displays no symmetry at all and is multifractal.  The black-and-white inset shows
the cross-sectional density in the $2 \times 2$ central region of the phase-plane section. $T = 1
+\epsilon \tanh(q)$ .
}
\end{figure}

On the microscopic side our Lyapunov spectra project with Harald led us to dozens of
studies of the effect of nonequilibrium conditions on the spectrum\cite{b19}. ``Color''
conductivity, where half the particles are accelerated to the right and the rest
to the left, provided a particularly clear example\cite{b19}.  Harald and Bill
pointed out the simple shift of the Lyapunov spectrum in 1987\cite{b19,b20}.  The
Lyapunov spectrum of
instability exponents shifts to more negative values with the sum of all exponents
equal to the rate at which entropy is extracted from the system by deterministic
thermostat forces.  Similar results were obtained for a variety of shear flows and
heat flows in systems with a variety of boundary conditions.

In {\it every} case nonequilibrium steady states occupy zero-volume fractal attractors
in phase space ( meaning that the total volume of the cells in phase space containing
any given fraction of the measure goes to zero as the small-cell limit is approached ).
This fact demonstrates the extreme rarity of nonequilibrium states.  The time
reversibility of the motion equations shows that the dynamically unstable repellor is
likewise of zero volume. The symmetry breaking revealed by these simulations shows
that only those flows satisfying the Second Law of Thermodynamics are observable\cite{b20}.

Exactly the same mechanism for irreversibility from reversible motion equations can be
seen in small systems. Even a single harmonic oscillator, exposed to a variable
temperature $T(q) = 1 + \epsilon\tanh(q)$ can generate a chaotic heat
flow and concentrate its stationary distribution on a fractal strange attractor.
The ``0532 Model'', which Bill mentioned\cite{b21}, incorporates ``weak'' control of
$\langle \ (p^2/T) \ \rangle$ and $\langle \ (p^4/T^2) \ \rangle$ and provides an
ergodic canonical distribution for the harmonic oscillator, enforced by a single
thermostat variable $\zeta$ :
$$
\dot q = p \ ; \ \dot p = -q -\zeta[ \ 0.05p + 0.32(p^3/T) \ ] \ ;
\ \dot \zeta = 0.05[ \ (p^2/T) - 1 \ ] + 0.32[ \ (p^4/T^2) - 3(p^2/T) \ ] \ .
$$
Bill showed the stationary phase-space cross section for $( \ \epsilon = 0 \ )$ .
Here  {\bf Figure 9} shows the flux through the $( \ \zeta = 0  \ )$ phase-space plane for
both the equilibrium flow and a nonequilibrium $( \ \epsilon = 0.5 \ )$ flow. The fractal
character of the nonequilibrium section is clear.  Notice also the symmetry breaking.
Time reversal breaks the symmetry of the local Lyapunov exponent ( indicated by color )
as well as the mirror symmetry present at equilibrium but absent at nonequilibrium.
In the nonequilibrium case the phase flow is {\it from} a repellor identical to the
attractor except for the signs of $p$ and $\zeta$ . It is also the case that the rate of entropy
production for the thermostated oscillator is equal to the rate of shrinkage
of the phase volume $\otimes$ :
$$
(\dot S/k) = \zeta[ \ 0.05(p^2/T) + 0.32(p^4/T^2) \ ] \ ; \
(\dot \otimes/\otimes) = -\zeta[ \ 0.05 + 0.96(p^2/T) \ ] 
$$
This is because the time average $\langle \ \zeta \dot \zeta \ \rangle \equiv 0$ implies that
$$
\langle \ 0.05\zeta[ \ (p^2/T) - 1 \ ] +
0.32\zeta[ \ (p^4/T^2) - 3(p^2/T) \ ]  \rangle = 0 \ ,
$$
or
$$
\langle \ \zeta[ \ 0.05(p^2/T) + 0.32(p^4/T^2) \ ] \ \rangle  =
\langle \ \zeta[ \ 0.05 + 0.96(p^2/T) \ ] \ \rangle \ , 
$$
so that the entropy production measured by the external thermostat is, when time averaged,
precisely equal to the loss rate of Gibbs' entropy in the nonequilibrium steady state.

Liouville's Theorem turned out to be the most useful tool in understanding these
ideas\cite{b9}. This  flow equation for thermostated systems describes the phase-volume
loss in terms of the friction coefficients imposing nonequilibrium constraints and in terms
of the Lyapunov spectrum. Stationary flows satisfying the Second Law of Thermodynamics
correspond to strange attractors, with shrinking phase volume.  Flows violating the
Law are both hard to find ( in the sense that they occupy zero phase volume ) and
dynamically unstable ( in the sense that they have a positive unstable Lyapunov sum )
 providing a mechanical analog for the Second Law of Thermodynamics directly from a
determinstic time-reversible mechanics.  Let us look at one particular manybody case,
the $\phi^4$ model, which is both simple and profound.

\subsection{The $\phi^4$ Model for Chaos and Fourier Heat Conduction}

\begin{figure}
\includegraphics[width=4in,angle=-90]{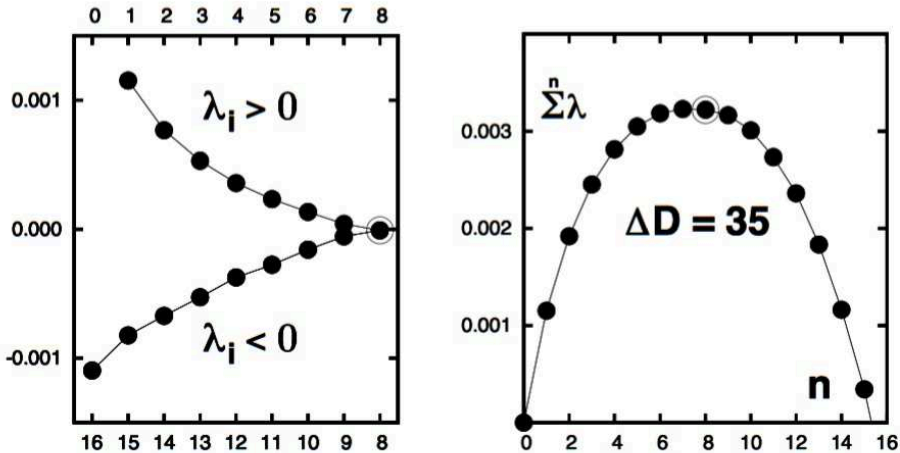}
\caption{
Lyapunov exponents ( left ) and their sums (right ), for a 24-particle $\phi^4$ chain with 
``cold'' temperature $0.003$ and ``hot'' temperature $0.027$ .  The dimensionality of the phase-space
attractor is only 15 while the dimensionality of the entire phase space is 50.
}
\end{figure}

Aoki and Kusnezov emphasized the utility of the $\phi^4$ model for statistical mechanics
and explored its chaos and heat conductivity.  It is a model of simplicity\cite{b23,b24}. 
Nearest-neighbor particles are joined by Hooke's-Law springs. Each particle is also
tethered to its lattice site with a quartic tether.  The $\phi^4$
model Hamiltonian has kinetic, tether, and harmonic contributions :
$$
{\cal H} = {\cal H}_K + {\cal H}_t + {\cal H}_H =
\sum_i [ \ (p_i^2/2) + (q_i^4/4) \ ] + \sum_{i<j} (q_{ij}^2/2) \ .
$$

At low temperature there is no heat conductivity as energy can travel through
a harmonic chain at the speed of sound.  At high temperature there is no heat conductivity
because the chain behaves as a system of independent quartic oscillators.  But it turns out
that the useful range of energies where the chain exhibits Fourier conductivity extends
over about nine useful orders of magnitude in between ! 
 
In 2002 Harald and Bill studied the family of two-dimensional $\phi^4$ crystals ranging
in size from $4\times 4$ to $12 \times 12$ with boundary temperatures of 0.001 and 0.009.
A fit to the data established that the dimensionality loss approaches 43 in the
large-system limit while the phase space coordinates associated with thermostating are
5 for each thermostated particle\cite{b25}.

The one-dimensional case is more dramatic. In our most recent book we provided two more
examples of largescale dimensionality loss using temperatures of 0.003 and 0.027. We
found that the dimensionality of a 24-particle attractor in its 50-dimensional phase
space was about 15. See {\bf Figure 10 }. A 36-particle attractor in its 74-dimensional phase space had an
attractor dimensionality of about 30.  By 2005 we had a good understanding of many
such simple models that illustrated the correspondence between microscopic fractal Lyapunov-unstable
systems and phenomenological macroscopic fluid mechanics.

\section{Our Ongoing Working Vacation in Ruby Valley Nevada}
\begin{figure}
\includegraphics[width=4.5in,angle=-90]{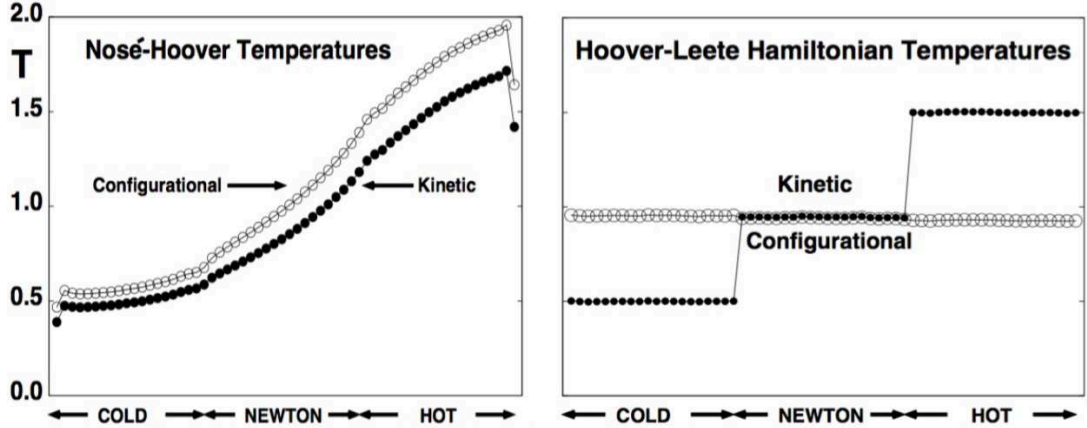}
\caption{
Kinetic and configurational temperature profiles for a 60-particle $\phi^4$ chain with
the configurational temperatures shown as open circles.
The temperatures on the left are solutions for the Nos\'e-Hoover equations of motion.
The particle temperatures are averages over a billion-timestep simulation with $dt = 0.01$
using a fourth-order Runge-Kutta integrator.
Simulations with longer chains are in agreement with Aoki and Kusnezov's work \cite{b23,b24}.
The temperatures on the right correspond to using the Hoover-Leete Lagrangian/Hamiltonian thermostat
to maintain the kinetic energies of the 20-particle cold and hot regions.  Despite the successful
thermostating of the reservoirs there is {\it no heat flow}.  The Hoover-Leete configurational
temperatures are nearly constant, and are equilibrated with the 20 Newtonian particles in the center
of the $\phi^4$ chain.  This simulation included a billion timesteps with $dt = 0.001$ .
}
\end{figure}

Bill spent much of my last ``work'' year in Livermore ( 2005 ) building us a new home in
Ruby Valley just a couple of miles from the crest of the Ruby Mountains.  Ruby Valley is 
a close-knit ranching community in northeastern Nevada.  The cows, mostly Aberdeen Angus,
outnumber the people by orders of magnitude and the telephone ``book'' is a single page,
distributed annually at Christmas along with copies of a Christmas Card from each valley
Family.  Communication is vital to research.  With airplanes and the internet we have
continued our active research life in Ruby Valley.  The cooperative and supportive nature
of science is clear to us as we number our collaorators at well over 100 with at least
that many more colleagues.  Heat flow, ergodicity, chaos, and shockwaves have been our
main research activities in Ruby Valley, along with related trips to Austria, Australia,
England, Japan, Mexico, Poland, Russia, and Spain\cite{b26,b27}.

\subsection{Hamiltonian Thermostats Do Not Promote Heat Flow}

One of our earliest findings in Ruby Valley illustrates a shortcoming of Hamiltonian
mechanics. It cannot describe steady heat flow.  We happened on this finding in 2007\cite{b28}
and revisited it in 2013\cite{b29} in response to unjustified claims for a ``Hamiltonian
thermostat'' which turned out to be misleading and worthless.  We have favored deterministic
thermostating and the kinetic definition of temperature ever since Bill Ashurst's work
in the early 1970s.  We thought it would be instructive to compare heat transport with the
simple Nos\'e-Hoover thermostat to heat transport with straightforward Hamiltonian thermostats.

We considered three varieties of Hamiltonian thermostats\cite{b30}, Nos\'e's, which reproduces
the entire canonical distribution, and two Lagrange-multiplier approaches leading to Hamiltonians
which constrain the kinetic or the configurational temperatures of selected degrees of freedom.  We found
generally that {\it Hamiltonian mechanics is incompatible with heat flow}.  If a cold thermostat extracts
heat while a hot thermostat provides it, both at the same rate $\dot Q$ the system's steady-state
entropy change is {\it negative} ! :
$$
\dot S = - (\dot Q/T_{\rm cold}) + (\dot Q/T_{\rm hot}) \simeq -(\dot Q\Delta T/T^2) < 0 \ .
$$
Here $\dot Q > 0$ is the ( time-averaged ) rate at which heat flows both in and out
of the system. Gibbs' relation linking entropy to phase volume $S = k\ln(\otimes)$ suggests that
any steady heat flow implies a loss of phase volume and is incompatible with Liouville's Theorem.
So the question is : What actually happens if the two heat reservoirs are modelled by Hamiltonian
mechanics ?  To find out we tried connecting several types of Hamiltonian thermostats to a
Newtonian system in a sandwich fashion.  The thermostats considered included : [ 1 ] Nos\'e's
original Hamiltonian\cite{b7,b8}, [ 2 ] the ``Hoover-Leete'' Hamiltonian\cite{b30} which constrains
the kinetic energy to a fixed constant :
$$
{\cal H}_{HL} = 2\sqrt{K(p)K(\dot q)} - K(\dot q) + \Phi \longrightarrow
\dot q = p\sqrt{K(\dot q)/K(p)} \longrightarrow
K(\dot q) \ {\rm constant} \ ,
$$ 
and [ 3 ] a Hamiltonian which constrains the configurational temperature to a fixed constant.  The
configurational temperature\cite{b31} folows from an integration by parts in the canonical ensemble:
$$
\int (\nabla {\cal H})^2e^{-{\cal H}/kT}dq = kT\int \nabla^2 {\cal H}e^{-{\cal H}/kT}dq \rightarrow
kT = \frac{\langle \ (\nabla {\cal H})^2 \ \rangle}{\langle \ \nabla^2{\cal H} \ \rangle} \ .
$$

An extremely interesting thing happens when any one of these three Hamiltonian thermostats is applied to the
$\phi^4$ heat-transfer problem. {\it There is no heat flow} !  {\bf Figure 11} illustrates a typical case,
with the thermostated kinetic temperature in the hot and cold regions having the specified cold and hot
values but with no current at all flowing through the Newtonian regions separating the two reservoirs.
These problems make the point that any {\it non}equilibrium simulation of heat flow with molecular dynamics
is necessarily intrinsically {\it non}Hamiltonian. This conclusion is surprising, as was also the fractal
nature of nonequilibrium distribution functions which came to light some thirty years ago in low-dimensional
simulations of nonequilibrium transport in phase spaces of three or four space dimensions.

\subsection{Shockwaves Revisited}

\begin{figure}
\includegraphics[width=4.5in]{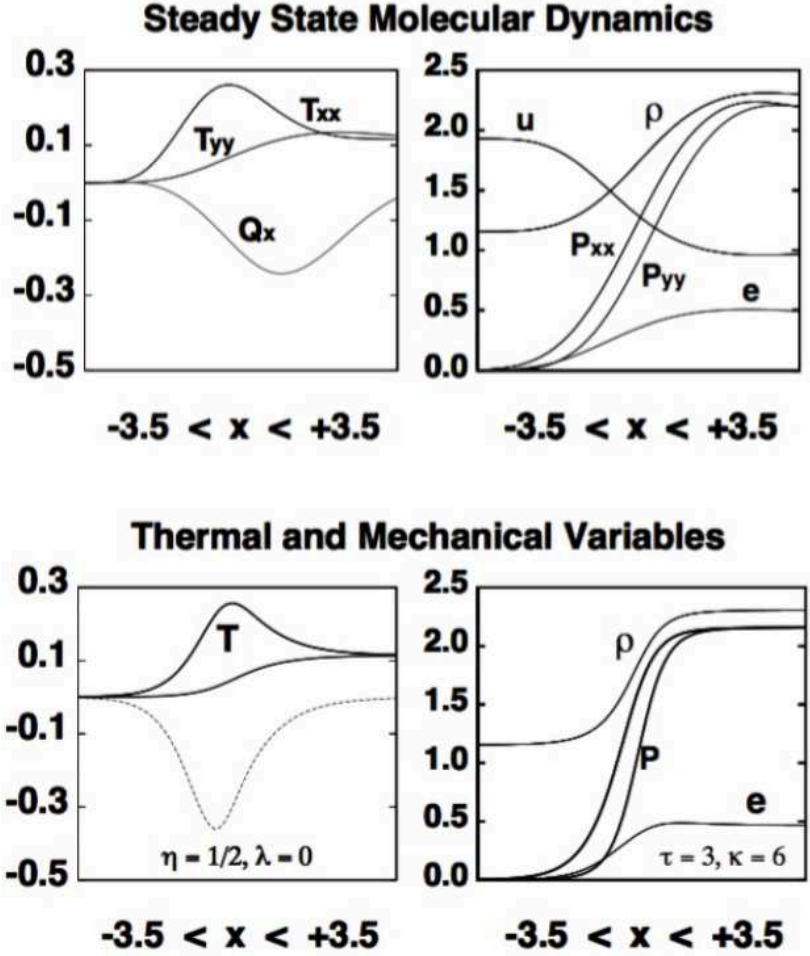}
\caption{
Molecular dynamics profiles (top) of mechanical and thermal variables generated
with Lucy's one-dimensional weight function $w(r < h = 3) = (5/12)[ \ 1 + r \ ][ \ 1 - (r/3) \ ]^3$ .
Mechanical and thermal variables (bottom) for a one-dimensional shockwave using the 
triangular-lattice ``cold'' curve, an allocation of the work and heat to $T_{xx}$ ,
and a tensor thermal conductivity with a thermal relaxation time of 3 .  For details see Reference 27.
}
\end{figure}

\begin{figure}
\includegraphics[width=4.5in,angle=-90]{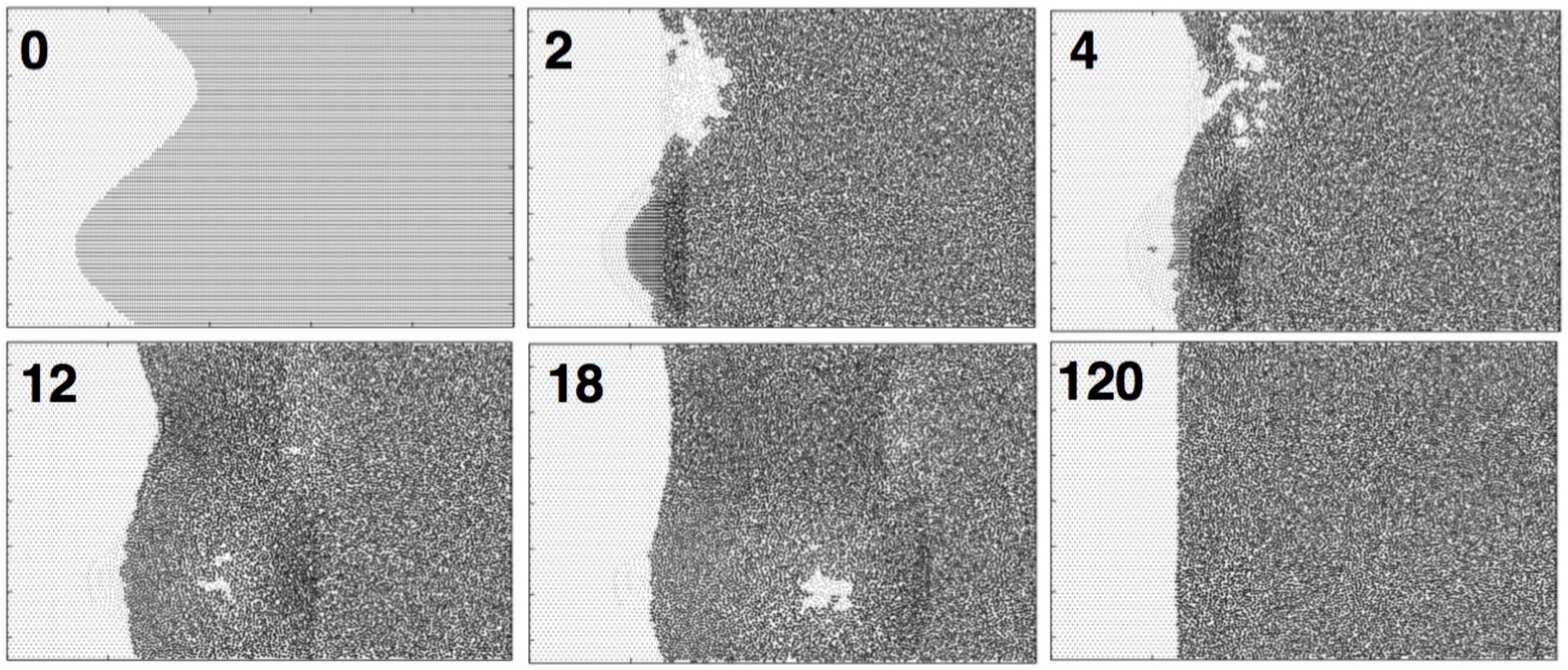}
\caption{
Evolution of a sinusoidal shockwave toward planarity.  Snapshots distinguish particles with lower
and higher densities than average. The window shown has length 100 with a periodic vertical height of
$40\sqrt{3}$ .  ``Cold'' stress-free material at the left has the close-packed
triangular-lattice structure with a nearest-neighbor spacing of unity. ``Hot'' fluid at the right has
twice that density.  Particle motion is left-to-right with input speed twice the hot-fluid exit
speed.  Snapshot times appear in the upper left corners. Notice the low-density tensile waves in the
third, fourth, and fifth frames.  The pair potential is short-ranged : 
$\phi(r<1) = (10/\pi)(1-r)^3$ .  The window shown moves at the shock velocity.
}
\end{figure}

Bill's early work on shockwaves was published in the Fall of 1967 in a summary of work at the Livermore Laboratory
presented by Bill's group leader Russ Duff at a meeting on the ``Behavior of Dense Media Under High Dynamic
Pressure'' in Paris\cite{b32}. Ten years later Klimenko and Dremin published dynamical studies\cite{b33} of two dense-fluid shockwaves for a
model of argon.  Ten years after that Brad Holian and Galen Straub at Los Alamos collaborated with Bill and Bill
Moran to study a 400-kilobar shockwave in a model of argon, just below the pressure and temperature at which the
fluid ionizes\cite{b34}.  Thirty years later with Paco Uribe\cite{b26,b27} we carried out studies confronting a detailed model capable
of matching the physical effects observed with molecular dynamics : [ 1 ] a large disparity between the
longitudinal and transverse temperatures, [ 2 ] a time delay in the response of stress to strain and heat flux to
a thermal gradient, and [ 3 ] a partition of heat and work into separate longitudinal and transverse components.
These features are capable of describing the far-from-equilibrium states found in strong fluid shockwaves.  

In correlating the results of molecular dynamics simulations with continuum models the smooth-particle averaging
is invaluable.  With SPAM averaging it was possible to establish the stability of planar shockwaves in two-dimensional systems
as is shown in {\bf Figure 13}.  Throughout we have observed marked differences between the longitudinal and
transverse temperatures as well as a time delay between the strain rate and the shear stress and between the
temperature gradient(s) and the heat flux.  With Paco Uribe in Mexico we were able to formulate and implement a
detailed model capable of describing all of these effects.  It is interesting that unless the delay time is brief,
on the order of a collision time, the constitutive model becomes unstable.

\subsection{Thermostats and India}
In 2014 Bill received an interesting paper to review for The Journal of Chemical Physics\cite{b35}.  The
authors, Baidurya Bhattacharya and his student Puneet Patra had the clever idea of thermostating both the kinetic
and the configurational temperatures simultaneously.  Bill thought this was a fine idea and wrote to them that he
enjoyed their paper and was looking forward to its publication.  The rapid-fire emails that followed soon led to
the appearance of Baidurya in Ruby Valley for a visit, and then in 2015 to a visit with his Family, which makes
regular trips to the United States in the summers. This led to a very productive collaboration and friendship.

We'll be traveling to Kharagpur in December for some lectures at Professor Bhattacharya's Institute, which we hope
will turn out to be yet another book. This meeting led, though indirectly, to solving a problem that had puzzled
Bill for 30 years : how to thermostat an ergodic canonical  harmonic oscillator with a single thermostat variable.
The Patra-Bhattacharya idea led first to simultaneous weak control of coordinates and momenta and finally to weak
control of two kinetic-energy moments. Several specific examples proved ergodicity for the oscillator and the
simple pendulum, but not ( so far ) for the quartic potential.  This same work also led us into collaborations with
Clint Sprott ( University of Wisconsin-Madison ), a wonderful colleague who has furnished many stimulating ideas as
well as his graphical expertise which is illustrated here in {\bf Figure 9}.

\subsection{Moral}

We are grateful to you all for your support, collaborations, friendship, and love over the years.  Thanks for your
attention to our reminiscences. Bill specially wants to thank Berni Alder for his participation here and for 
his extending a helping hand in so
many ways that were crucial to the good life Bill and Carol have enjoyed so far. We urge those of you who are
younger to reflect on your good fortune in being a part of our progress in understanding the world around us.  
Finally, we offer our thanks to Karl Travis and Fernando Bresme for their thoughtfulness and skill in organizing
this meeting at Sheffield: ``Advances in Theory and Simulation of Nonequilibrium Systems''.


\begin{thebibliography}{99}

\bibitem{b1}  W. G. Hoover, C. G. Hoover, and H. A. Posch, ``Lyapunov Instability of Pendula, Chains
              and Strings'', Physical Review A {\bf 41}, 2999-3004 (1990).

\bibitem{b2}  W. G. Hoover, {\it Computational Statistical Mechanics} (Elsevier, Amsterdam, 1991).

\bibitem{b3}  W. G. Hoover, A. J. De Groot, C. G. Hoover, I. F. Stowers, T. Kawai, B. L. Holian,
              T. Boku, S. Ihara, and J. Belak, ``Large-Scale Elastic-Plastic Indentation Simulations
              {\it via} Molecular Dynamics'', Physical Review A {\bf 42}, 5844-5853 (1990).

\bibitem{b4}  J. S. Kallman, W. G. Hoover, C. G. Hoover, A. J. De Groot, S. Lee, and F. Wooten,
              ``Molecular Dynamics of Silicon Indentation'', Physical Review B {\bf 47},
              7705-7709 (1993).

\bibitem{b5}  J. S. Kallman, A. J. De Groot, C. G. Hoover, W. G. Hoover, S. M. Lee, and F. Wooten,
              ``Visualization Techniques for Molecular Dynamics'', IEEE Computer Graphics and Applications
              {\bf 15}, 72-77 (November, 1995).

\bibitem{b6}  H. J. M. Hanley, {\it Nonlinear Fluid Behavior}, Proceedings of a 1982 Conference in
              Boulder, Colorado, published as Physica {\bf 118A}, 1-454 (1983).

\bibitem{b7}  S. Nos\'e, ``A Molecular Dynamics Method for Simulations in the Canonical Ensemble'',
              Molecular Physics {\bf 52}, 255-268 (1984).

\bibitem{b8}  S. Nos\'e, ``A Unified Formulation of the Constant Temperature Molecular Dynamics'',
              The Journal of Chemical Physics {\bf 81}, 511-519 (1984).

\bibitem{b9}  W. G. Hoover, ``Canonical Dynamics: Equilibrium Phase-Space Distributions'', Physical
              Review A {\bf 31}, 1695-1697 (1985).

\bibitem{b10} G. Ciccotti and W. G. Hoover, {\it Molecular Dynamics  Simulations of Statistical Mechanical
              Systems}, Proceedings of the 1985 Enrico Fermi International School of Physics at Varenna
              (Elsevier, New York, 1986), 622 pages.

\bibitem{b11} B. L. Holian, W. G. Hoover, and H. A. Posch, ``Second-Law Irreversibility of Reversible
              Mechanical Systems''  =  ``Resolution of Loschmidt's Paradox: the Origin of Irreversible
              Behavior in Reversible Atomistic Dynamics'', Physical Review Letters {\bf 59}, 10-13 (1987).

\bibitem{b12} L. B. Lucy, ``A Numerical Approach to the Testing of the Fission Hypothesis'', \
              Astronomical Journal {\bf82}, 1013-1024 (1977).

\bibitem{b13} J. J. Monaghan, ``Smoothed Particle Hydrodynamics'', Annual Review of Astronomy
              and Astrophysics {\bf 30}, 543-574 (1992).

\bibitem{b14} Wm. G. Hoover and H. A. Posch, ``Entropy Increase in Confined Free Expansions {\it via}
              Molecular Dynamics and Smooth-Particle Applied Mechanics'', Physical Review E {\bf 59},
              1770-1776 (1999).

\bibitem{b15} Wm. G. Hoover, H. A. Posch, V. M. Castillo, and C. G. Hoover, ``Computer Simulation of
              Irreversible Expansions {\it via} Molecular Dynamics, Smooth Particle Applied Mechanics,
              Eulerian, and Lagrangian Continuum Mechanics'', Journal of Statistical Physics {\bf 100},
              313-326  (2000).

\bibitem{b16} O. Kum, W. G. Hoover, and H. A. Posch, ``Viscous Conducting Flows with
              Smooth-Particle Applied Mechanics'', Physical Review E {\bf 52}, 4899-4908 (1995).

\bibitem{b17} V. M. Castillo, Wm. G. Hoover, and C. G. Hoover, ``Coexisting Attractors in
              Compressible Rayleigh-B\'enard Flow'', Physical Review E {\bf 55}, 5546-5550 (1997).

\bibitem{b18} V. M. Castillo and Wm. G. Hoover, ``Entropy Production and Lyapunov Instability at
              the Onset of Turbulent Convection'', Physical Review E {\bf 58}, 7350-7354 (1998).

\bibitem{b19} W. G. Hoover and H. A. Posch, ``Direct Measurement of Equilibrium and Nonequilibrium
              Lyapunov Spectra'', Physics Letters A {\bf 123}, 227-230 (1987).

\bibitem{b20} H. A. Posch and W. G. Hoover, ``Chaotic Dynamics in Dense Fluids'', {\it Liquids of Small
              Molecules}, Proceedings of a Conference at Santa Trada, Calabria, Italy, presented on
              22 September 1987 and available in the book of abstracts published by the European Physical
              Society.

\bibitem{b21}  P. K. Patra, J. C. Sprott, W. G. Hoover and C. G. Hoover, ``Deterministic Time-Reversible
              Thermostats : Chaos, Ergodicity, and the Zeroth Law of Thermodynamics'', Molecular
              Physics {\bf 113}, 2863-2872 (2015).

\bibitem{b22} K. Aoki and D. Kusnezov, ``Bulk Properties of Anharmonic Chains in Strong
              Thermal Gradients: Nonequilibrium $\phi^4$ Theory'', Physics Letters A
              {\bf 265}, 250-256 (2000).

\bibitem{b23} K. Aoki and D. Kusnezov, ``Nonequilibrium Steady States and Transport
              in the Classical Lattice $\phi^4$ Theory'', Physics Letters B {\bf 477},
              348-354 (2000).

\bibitem{b24} H. A. Posch and W. G. Hoover, ``Large-System Phase-Space Dimensionality Loss
              in Stationary Heat Flows'', Physica D {\bf 187}, 281-293 (2004).

\bibitem{b25} W. G. Hoover and C. G. Hoover, {\it Simulation and Control of Chaotic Nonequilibrium
              Systems} (World Scientific, Singapore, 2015).

\bibitem{b26} W. G. Hoover, C. G. Hoover, and F. J. Uribe, ``Flexible Macroscopic Models for Dense-Fluid
              Shockwaves: Partitioning Heat and Work; Delaying Stress and Heat Flux; Two-Temperature
              Thermal Relaxation'', Proceedings of the International Summer School Conference:
              ``Advanced Problems in Mechanics-2010'' organized by the Institute for Problems in
              Mechanical Engineering of the Russian Academy of Sciences in Mechanics and Engineering
              under the patronage of the Russian Academy of Sciences = arXiv 1005.1525 (2010).

\bibitem{b27} F. J. Uribe, W. G. Hoover, C. G. Hoover, ``Maxwell and Cattaneo's Time-Delay Ideas Applied
              to Shockwaves and the Rayleigh-B\'enard Problem'', Computational Methods in Science and
              Technology {\bf 19}, 5-12 (2013).

\bibitem{b28} Wm. G. Hoover and C. G. Hoover, ``Hamiltonian Dynamics of Thermostated Systems:
              Two-Temperature Heat-Conducting $\phi^4$ Chains'', Journal of Chemical Physics
              {\bf 126}, 164113 (2007).

\bibitem{b29} W. G. Hoover and C. G. Hoover, ``Hamiltonian Thermostats Fail to Promote Heat Flow'',
              Communications in Nonlinear Science and Numerical Simulation {\bf 18}, 3365-3372
              (2013).

\bibitem{b30}  T. Leete, {\it The Hamiltonian Dynamics of Constrained Lagrangian Systems} (Master's
              Thesis, West Virginia University, 1979).

\bibitem{b31} K. P. Travis and C. Braga, ``Configurational Temperature Control for Atomic and Molecular
              Systems'', The Journal of Chemical Physics {\bf 128}, 014111 (2008) = arXiv 0709.1575.

\bibitem{b32} R. E. Duff, W. H. Gust, E. B. Royce, M. Ross, A. C. Mitchell, R. N. Keeler, and W. G.
              Hoover, ``Shockwave Studies in Condensed Media'', in {\it Behavior of Dense Media Under
              High Dynamic Pressures} (Gordon and Breach, New York, 1968).

\bibitem{b33} V. Y. Klimenko and A. N. Dremin, ``Structure of Shockwave Front in a Liquid'' in
              {\it Detonation}, Chernogolovka, edited by O. N. Breusov {\it et alii} (Akademiya Nauk,
              Moscow, SSSR, 1978), pages 79-83.

\bibitem{b34} B. L. Holian, W. G. Hoover, B. Moran, and G. K. Straub, ``Shockwave Structure {\it via}
              Nonequilibrium Molecular Dynamics and Navier-Stokes Continuum Mechanics'', Physical
              Review A {\bf 22}, 2798-2808 (1980).

\bibitem{b35} P. K. Patra and B. Bhattacharya, ``A Deterministic Thermostat for Controlling Temperature
              Using All Degrees of Freedom'', The Journal of Chemical Physics {\bf 140}, 064106 (2014).

\end{thebibliography}
\end{document}